\documentclass[twocolumn,superscriptaddress,showpacs,preprintnumbers,amsmath,amssymb,prx]{revtex4}

\usepackage{graphicx}
\usepackage{dcolumn}
\usepackage{bm}
\usepackage{amsmath,amssymb}
\usepackage{color}

\begin{document}
\title{Fulde-Ferrell-Larkin-Ovchinnikov states in a superconducting ring with 
magnetic fields: Phase diagram and the first-order phase transitions
}
\author{Ryosuke\ Yoshii}
\thanks{Present address: Laboratoire de Physique des Solides, CNRS
UMR-8502, Univ. Paris Sud, 91405 Orsay Cedex, France.}
\affiliation{Department of Physics, Osaka university, Machikaneyama, Toyonaka, Osaka 560-0043, Japan}
\affiliation{Yukawa Institute for Theoretical Physics, Kyoto University, Kitashirakawa Oiwake-Cho, Kyoto 606-8502, Japan}
\affiliation{Research and Education Center for Natural Sciences, Keio University, 4-1-1 Hiyoshi, Kanagawa 223-8521, Japan}

\author{Satoshi\ Takada}
\affiliation{Yukawa Institute for Theoretical Physics, Kyoto University, Kitashirakawa Oiwake-Cho, Kyoto 606-8502, Japan}

\author{Shunji\ Tsuchiya}
\affiliation{Center for General Education, Tohoku Institute of Technology,
35-1 Yagiyama Kasumi-cho, Taihaku-ku, Sendai, Miyagi 982-8577, Japan}
\affiliation{Research and Education Center for Natural Sciences, Keio University, 4-1-1 Hiyoshi, Kanagawa 223-8521, Japan}

\author{Giacomo\ Marmorini}
\affiliation{Yukawa Institute for Theoretical Physics, Kyoto University, Kitashirakawa Oiwake-Cho, Kyoto 606-8502, Japan}
\affiliation{Research and Education Center for Natural Sciences, Keio University, 4-1-1 Hiyoshi, Kanagawa 223-8521, Japan}

\author{Hisao\ Hayakawa}
\affiliation{Yukawa Institute for Theoretical Physics, Kyoto University, Kitashirakawa Oiwake-Cho, Kyoto 606-8502, Japan}
\date{\today}

\author{Muneto\ Nitta}
\affiliation{Department of Physics, Keio University, 4-1-1 Hiyoshi, Kanagawa 223-8521, Japan
}
\affiliation{Research and Education Center for Natural Sciences, Keio University, 4-1-1 Hiyoshi, Kanagawa 223-8521, Japan}

\begin{abstract}
We find the angular Fulde-Ferrell-Larkin-Ovchinnikov (FFLO) 
states (or the twisted kink crystals) in which a phase and an amplitude of a pair potential modulate simultaneously in a quasi-one-dimensional superconducting ring with a static Zeeman magnetic field applied on the ring and static Aharonov-Bohm magnetic flux penetrating the ring. 
The superconducting ring with magnetic flux produces a persistent current, 
whereas the Zeeman split of Fermi energy results in the spatial modulation of the pair potential. 
We show that these two magnetic fields stabilize 
the FFLO phase in a large parameter region of the magnetic fields. 
We further draw the phase diagram with the two kinds of first-order phase transitions; 
one corresponds to phase slips separating the Aharonov-Bohm magnetic flux, 
and the other separates the number of peaks of the pair amplitude for the Zeeman magnetic field.

\end{abstract}

\pacs{74.81.-g, 03.75.Lm, 74.78.-w, 74.25.Dw}

\maketitle
\section{Introduction} 
Superconductivity is one of the most exotic states of matter, appearing in a broad range of systems in nature, 
from metallic superconductors to neutron stars. 
Superconducting states stem from a condensation of Cooper pairs, 
which are made of fermions with two different inner states with opposite momentum near each Fermi surface. 
The different forms of the Cooper pairs appear in various physical systems, 
such as those with different spins of electrons for metallic superconductors, 
those with different chirality of quarks for a dynamical mass generation 
in QCD, and those with different atomic states for superfluids in ultracold Fermi gases. 
After the BCS theory was proposed, 
an exotic state called a Fulde-Ferrell-Larkin-Ovchinnikov (FFLO) state 
was conjectured, in which the pair potential has spatial modulation \cite{FF,LO}. 
When a population imbalance exists between those different inner states, 
the difference of the sizes of the Fermi surfaces results in 
a finite total momentum of the Cooper pairs, yielding the FFLO state.  
The modulation of the pair potentials can be classified into two classes: 
a phase modulation which is called the Fulde-Ferrell (FF) state \cite{FF} 
and an amplitude modulation which is called the Larkin-Ovchinnikov (LO) state \cite{LO}.  
When a persistent current exists, 
the pair potential becomes a plane wave like $\Delta \propto e^{i m x}$ 
with a constant $m$, resulting in a FF-like state. 
The population imbalance induced, for example, 
by a magnetic field on a quasi-one-dimensional (quasi-1D) superconductor  results in a LO state, 
described by a sine like shape, $\Delta \propto \mathrm{sn} (x, \nu)$, 
with the elliptic parameter $\nu$ \cite{Machida}. 
The FFLO states have been mostly analyzed thus far in the basis of the Ginzburg-Landau (GL) equation in various systems, 
e.g., the superconductor in magnetic field \cite{Klein, Zyuzin, Aoyama}, superconductor-ferromagnet heterostructures \cite{Buzdin3, Bergeret}.  
It is, however, known that the GL equation is only valid in the vicinity of the critical temperature. 
Thus the GL formalism is not appropriate to discuss the lowest energy state at temperatures much lower than the critical temperature, 
where one has to use the Bogoliubov-de Gennes (BdG) formalism. 
Recent developments of research in cold atomic Fermi gases have renewed interest in the FFLO state \cite{Radzihovsky}, 
and observation of a spin-polarized superfluid state was reported \cite{Liao} in which it is expected that the FFLO state has been achieved. 
However, direct observation of its oscillating order parameter is still lacking.
Besides metallic superconductors and cold Fermi gases, 
these states have been attracting much attention in QCD,  
because they are also expected to appear in chiral condensates or diquark condensates of quarks 
at high density and/or high temperature \cite{Casalbuoni, Anglani}. 
However, in condensed matter systems, the direct confirmation of these states 
has not yet been achieved for 50 years since its proposal,  in spite of tremendous efforts \cite{Radovan, Yonezawa, Wu}.

As a simple setup to realize phase and/or amplitude modulations, we consider a superconducting ring.
When a ring is penetrated by a magnetic flux, the phase of the wave function on the ring depends on the magnetic flux even if magnetic field itself is not applied on the ring. 
This effect is known as the Aharonov-Bohm (AB) effect and can be used to make a persistent current for superconductors fabricated on a ring. 
The resulting pair potential becomes that of a FF-like state \cite{FF}. 
When the population imbalance is induced, for example, 
by a magnetic field on the ring, excess particles which cannot make a Cooper pair appear. 
If we consider the pair potential as a background potential and focus on the energies of these excess particles,  
a normal state, the LO state, and the BCS state are favorable in this order. 
On the other hand, if we focus on the energies of Cooper pairs, 
the BCS state, the LO state, and the normal state are favorable in this order. 
Thus the LO state appears between the BCS state and the normal state when the magnetic field is increased \cite{Machida}. 
A question arises about the competition between 
the AB effect and the population imbalance for different spins. 
In the presence of these two magnetic fields, 
the phase transition between the FF and LO states was reported in Ref.\ \cite{Quan}. 
Another group suggested the existence of the half-vortex state in a similar setup of ultracold atomic gases \cite{Yoshida}. 

In this paper, we demonstrate that a novel phase, an angular FFLO state or 
the so-called twisted kink 
(complex kink or gray soliton) crystal, is stabilized in which both amplitude and phase of the pair  potential are spatially modulated along a superconducting ring with the AB magnetic flux penetrating the ring and the static Zeeman magnetic field on the ring. 
We draw the phase diagram as a function of both magnetic fields 
by using the BdG formalism valid at temperatures much lower than the critical temperature (including $T=0$) 
and find the twisted kink crystals to be the lowest energy states in a large region of the parameter space. 
We find the two kinds of first-order phase transitions; one corresponds to phase slips 
separating the Aharonov-Bohm magnetic flux and another separates the number of peaks 
of the pair amplitude for the Zeeman magnetic field. 
The twisted kink crystal in an infinite system was found in high-energy physics 
as a self-consistent solution of the Gross-Neveu model in 1+1 dimensions 
or equivalently, the BdG equation with the Andreev approximation \cite{Basar, Basar2}. 
However, only a phase modulation (the FF state) was found to appear in the phase diagram 
of the Gross-Neveu model in 1+1 dimensions \cite{Basar:2009fg}.  
Our work is a proposal to realize 
an FFLO state with both phase and amplitude of the pair potential modulated.

\section{Analytic solution of twisted kink crystal state}
\begin{figure}
\includegraphics[width=10pc]{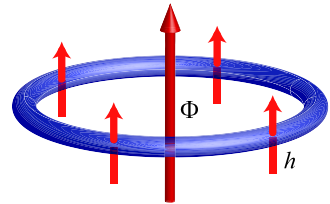}\hspace{1pc}
\caption{Schematic picture of our setup. The superconducting ring is penetrated by the magnetic flux $\Phi$, and the magnetic field $h$ is applied on the ring. }
\label{ring}
\end{figure}

In this section, we make a brief review of the analytical solution of twisted kink crystal state and 
we show that the method used can be generalized in the presence of the magnetic fields. 
We study a quasi-1D superconducting ring under magnetic fields (Fig.~\ref{ring})  by the mean-field BdG equation \cite{BdG}.  
Here, we assume that the radius of the ring is large enough compared to its width so that 
the curvature effect can be ignored. 
Although the mean-field approximation is not valid in strictly one dimension, 
we assume a quasi-1D system, which is more relevant to experiments and can be well described by the  BdG equation for quasiparticles $u(x)$ and $v(x)$ 
(we adopt the units $\hbar=1,~c=1,~e=1$): 
\begin{eqnarray}
&\left[
\begin{array}{cc}
H_{\uparrow}&\Delta(x)\\ 
\Delta^\ast(x) &-H_{\downarrow}^\ast
\end{array}
\right]
\left[
\begin{array}{c}
u(x) \\ v(x)
\end{array}
\right]
=E
\left[
\begin{array}{c}
u(x)\\v(x)
\end{array}
\right],\label{BdG0}\\
&  H_{\sigma}=\displaystyle\frac{1}{2M}\left[-i\frac{\partial}{\partial x}- \frac{\phi}{L}\right]^2-\mu_\sigma, 
\end{eqnarray}
where $\sigma$ ($=\uparrow,\downarrow$) stands for the spin and $M$ is the mass of the fermion. 
Here we have defined the $x$ coordinate along the ring and thus the system must be periodic in $x$.
We denote the length of the circumference by $L$.
The energy difference due to the Zeeman splitting, which stems from the magnetic fields $h$ applied on the ring, is included in the chemical potential for each spin state $\mu_\sigma=\mu-\sigma h$. 
Here $\mu$ is the chemical potential in the absence of the magnetic field. 
The effect from the AB flux penetrating the ring is introduced by the vector potential $\vec{A}=(A_x, A_y, A_z)$. 
For our system, the vector potential can be written as $\vec{A}=(\phi /L,~0,~0)$ by using the AB phase $\phi$.
Here the AB phase $\phi$ is defined as $\phi=2\pi\Phi/\Phi_0$, 
with the AB flux penetrating the ring $\Phi$ and flux quantum $\Phi_0=hc/e$. 
For simplicity, we restrict ourselves to the case $T=0$. 
In this case, the pair potential $\Delta(x)$ satisfies the gap equation
\begin{equation}
\Delta(x)=-2g^2\sum_{E_n<0}u_n(x)v_n(x)^*,
\label{Gap0}
\end{equation}
where $g$ is the attractive interaction between fermions with different 
spins and $n$ is the index for eigenstates. 

By using the transformation $\left[u(x),~v(x)\right]^T=e^{i\sigma_3\phi x/L}\left[\hat u(x),\  \hat{v}(x)\right]^T$ and $\Delta(x)=e^{2i\phi{x}/L}\hat \Delta(x)$ with Pauli's matrix $\sigma_3$, 
the AB flux dependence of Eqs.~(\ref{BdG0}) and (\ref{Gap0}) vanishes as 
\begin{eqnarray}
&\left[
\begin{array}{cc}
\hat H_{\uparrow}&\hat \Delta(x)\\ 
\hat \Delta^\ast(x) &-\hat H_{\downarrow}^\ast
\end{array}
\right]
\left[
\begin{array}{c}
\hat u(x) \\ \hat v(x)
\end{array}
\right]
=E
\left[
\begin{array}{c}
\hat u(x) \\ \hat v(x)
\end{array}
\right],\label{BdG_1}\\
& \hat H_{\sigma}=-\displaystyle\frac{1}{2M}\frac{\partial^2}{\partial x^2}-\mu_\sigma,
\end{eqnarray}
and
\begin{equation}
\hat \Delta(x)=-2g^2\sum_{E_n<0}\hat u_n(x)\hat v_n(x)^*.
\label{Gap_1}
\end{equation}
The effect of the AB phase appears only as a new boundary condition,
\begin{equation}
\hat \Delta(x+L)=e^{2i\phi }\hat \Delta(x).
\end{equation}

If the attractive interaction is smaller than the Fermi energy
$\varepsilon_{{\rm F}\sigma}=\mu_\sigma$, fermions near the Fermi surfaces form Cooper pairs. 
In this case, we may adopt the Andreev approximation \cite{Barsagi}. 
Let $\hat{u}(x)=e^{ik_{\mathrm{F}\uparrow}x} \hat{u}_0(x)$ and 
$\hat{v}(x)=e^{-ik_{\mathrm{F}\downarrow}x}\hat{v}_0(x)$, where $k_{\rm{F\sigma}}$ 
is the Fermi momentum $k_{\rm{F\sigma}}=\sqrt{2M\varepsilon_{\rm{F\sigma}}}$. 
Then, $\hat{u}_0(x)$ and $\hat{v}_0(x)$ vary much slower than the length scale of $1/k_{\mathrm{F}\sigma}$.
Neglecting the second derivative terms of $\hat{u}_0(x)$ and $\hat{v}_0(x)$, 
the BdG equation reduces to 
\begin{eqnarray}
&\left[
\begin{array}{cc}
-i v_{\mathrm{F}\uparrow}\frac{\partial}{\partial x}&\hat\Delta_0(x)\\ 
\hat\Delta_0^\ast(x) & i v_{\mathrm{F}\downarrow}\frac{\partial}{\partial x}
\end{array}
\right]
\left[
\begin{array}{c}
\hat u_0(x) \\ \hat v_0(x)
\end{array}
\right]
\simeq E
\left[
\begin{array}{c}
\hat u_0(x) \\ \hat v_0 (x)
\end{array}
\right],
\label{BdG}
\end{eqnarray}
where $v_{\mathrm{F}\sigma}=k_{\mathrm{F}\sigma}/M$ is the Fermi velocity and
$\hat{\Delta}_0=e^{-i(k_{\mathrm{F}\uparrow}+k_{\mathrm{F}\downarrow})x}\hat\Delta$.

\begin{figure}
\includegraphics[width=17pc]{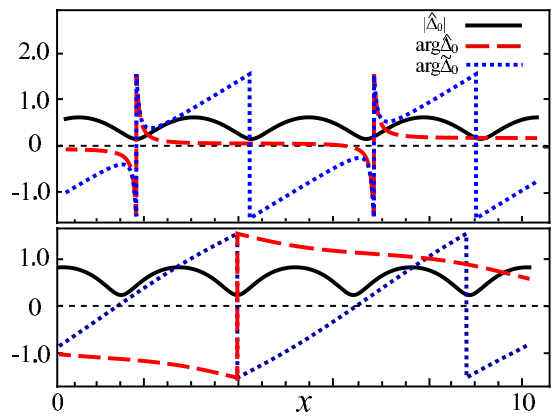}
\caption{The analytical solutions of FFLO phase for $\nu=0.5$ (upper figure) and $\nu=0.2$ (lower figure),  where $m=1$ and $\theta=4$ in both cases.
We plot the spatial profile for the absolute value (solid line) and the phase (broken line) of the pair potentials. 
We also plot the phase for $\tilde{\Delta}_0=\hat{\Delta}_0\times\exp(i\gamma{x})$ (dotted line), 
where we chose the smallest $\gamma>0$ which satisfies $\tilde{\Delta}_0(x+4K/A)=\tilde{\Delta}_0(x)$. 
The phase of the pair potential is calculated as $\mathrm{arg}\hat{\Delta}_0\equiv\tan^{-1}(\mathrm{Im}\hat \Delta_0/\mathrm{Re}\hat{\Delta}_0)$, which has a range $(-\pi/2,\pi/2)$. The actual phase should be read by adding $\pi/2$ at discontinuous points. 
For instance, in the lower panel, $\pi/2$ is added to $\mathrm {arg}\hat{\Delta}_0$ in the region of $x>4$, so one can see that $\mathrm{arg}\hat{\Delta}_0$ monotonically decreases. }
\label{tc_analytical}
\end{figure}

This approximated BdG equation (\ref{BdG}) and the gap equation (\ref{Gap_1}) are used in Refs.\ \cite{YoshiiPRB, YoshiiJPSJ}, except for the boundary condition. 
Thus the method used there can be applied to the present problem. 
It is known that the general solution for  the gap function $\hat{\Delta}_0(x)$ is 
\begin{align}
\hat \Delta_0(x)=
&-\alpha A 
\frac{\sigma(Ax+i{K}^\prime-i\theta/2)}
{\sigma(Ax+i{K}^\prime)\sigma(i\theta/2)}\nonumber\\
&\times\exp \left\{iAx(-i\zeta(i\theta/2)+i\mathrm{ns}(i\theta/2))+i\theta\eta_3/2\right\},
\label{FFLO}
\end{align}
where $\sigma,~\zeta$, and $\mathrm{ns}=1/\mathrm{sn}$ are, respectively, 
the Weierstrass $\sigma$, $\zeta$ functions, and Jacobi elliptic functions, 
characterized by the elliptic parameter $\nu$ and the half periods  $\omega_1$ and $\omega_3$ for real and imaginary direction, respectively. 
We set the half-periods to $\omega_1=K$ and $\omega_3=iK^\prime$, 
with $K(\nu)=\int_0^{\pi/2}dt(1-\nu\sin^2 t )^{-1/2}$ and $K^\prime\equiv {K}(1-\nu)$ \cite{Basar, Basar2, YoshiiPRB, YoshiiJPSJ}. 
The constant $\eta_3$ is defined by $\zeta (iK^\prime)$. 
The parameter $A$ represents the scale of the condensate as  $A=-2im\mathrm{sc}(i\theta/4)\mathrm{nd}(i\theta/4)$. 
Here $\mathrm{sc}=\mathrm{sn}/\mathrm{cn}$ and $\mathrm{nd}=1/\mathrm{dn}$ are Jacobi elliptic functions, 
and $m,~\theta$ are related to the amplitude and the phase modulation, respectively. 
In addition, we have introduced the imbalance parameter as $\alpha={\sqrt{v_{\mathrm{F}\uparrow}v_{\mathrm{F}\downarrow}}}/{v_{\mathrm{F}}}$ ($0\leq\alpha\leq1$) with $v_{\mathrm{F}}=\left(v_{\mathrm{F}\uparrow}+v_{\mathrm{F}\downarrow}\right)/2$. 
This solution has periodicity $l=2K/A$ for the amplitude of the pair potential as 
\begin{equation}
\Delta(x+l)=e^{2i\xi}\Delta(x),
\end{equation}
where
\begin{equation}
\xi=K[-i\zeta(i\theta/2)+i\mathrm{ns}(i\theta/2)-\eta \theta/2K],
\end{equation}
with $\eta=\zeta(K)$ \cite{Basar, Basar2}. 
Furthermore, it is known that this solution includes several previously known solutions as special cases, such as  the constant condensation (BCS state), the FF state, the LO state, 
the complex (twisted) kink, and the real kink \cite{Shei}. 
We again note that the effect of AB flux enters via the uniform phase modulation 
$\Delta(x) =\exp ({2i\phi x}/L)\exp [{i(k_{\mathrm{F}\uparrow}+k_{\mathrm{F}\downarrow})x}]\hat{\Delta}_0(x)$.
We plot typical solutions for $\nu=0.7,~\theta=3$ and $\nu=0.9,~\theta=3$ in Fig.\ \ref{tc_analytical}. 
We also plot the phase for $\tilde{\Delta}_0=\hat{\Delta}_0\times\exp(i\gamma{x})$ (dotted line), 
where we chose the smallest $\gamma>0$ which satisfies $\tilde{\Delta}_0(x+4K/A)= \tilde{\Delta}_0(x)$. 
In our ring geometry, when $\gamma L$ is identical to $2\phi+(k_{\mathrm{F}\uparrow}+k_{\mathrm{F}\downarrow})L$ (mod $2\pi$), $\tilde{\Delta}_0(x)$ becomes the solution of the BdG equation (\ref{BdG_1}) and the gap equation (\ref{Gap_1}).
Another important point is that the amplitude of the condensate does  not vanish in the whole region for the twisted kink crystal state. 
The above solution has a possibility to be stabilized by the  two kinds of magnetic fields introduced above. 
In the following, we numerically calculate the pair potentials for BdG and gap equations and the corresponding free energies in the presence of magnetic fields. 

\section{Existence of the FFLO state} 
In order to show that the above FFLO phase is stabilized in the presence of the magnetic fields, 
we numerically calculate the free energy in our system for small finite temperature. 
In our numerical calculations, we discretize Eq.~(\ref{BdG0}) and solve it self-consistently together with Eq.~(\ref{Gap0}). 
The discretized BdG equation becomes
\begin{equation}
\sum_j\left[
\begin{array}{cc}
H_{i, j, \sigma}&\Delta_i \delta_{i, j}\\
\Delta_i^\ast\delta_{i, j}& -H^\ast_{i, j, \bar\sigma} 
\end{array}
\right]
\left[\begin{array}{c}
u^\nu_{j\sigma}\\
v^\nu_{j\bar\sigma}
\end{array}
\right]=E_\nu
\left[\begin{array}{c}
u^\nu_{i\sigma}\\
v^\nu_{i\bar\sigma}
\end{array}
\right],
\label{BdGN}
\end{equation}
where $H_{i,j,\sigma}=-t_{i,j}-\mu\delta_{i,j}+\sigma h\delta_{i,j}$, and $i$, $\sigma,~\nu$ label the site, 
the spin of the particle, and eigenenergy, respectively. 
We treat the effect of the AB flux penetrating the ring by using the Peierls phase, 
$t_{i, i+1}=t\exp (i \phi/N)$, $t_{i+1,i}=t\exp(-i\phi/N)$ with transfer integral $t$ and the total site number $N$, 
where we only consider the nearest-neighbor hopping. 
\begin{figure}
\includegraphics[width=16pc]{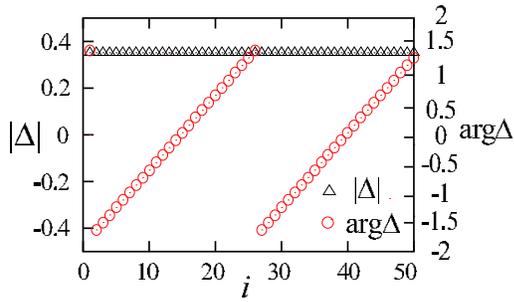}
\caption{Typical profile of the pair potential of the FF state ($\phi/2\pi=0.3,~h=0$). 
The amplitude of the pair potential is measured in the unit of the transfer integral. 
In this phase the amplitude of the pair potential is a constant and phase modulates.}
\label{FigFF}
\end{figure}
\begin{figure}
\includegraphics[width=12pc]{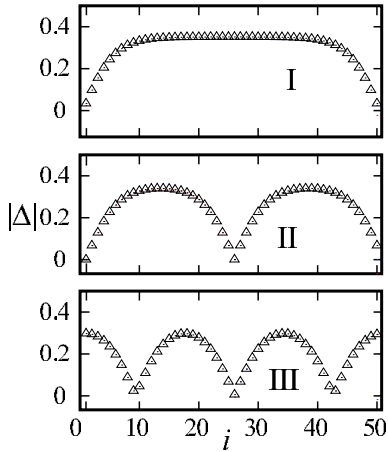}
\caption{Typical profiles of the pair potential of the LO (FFLO) states. The top, middle, and bottom figures are the absolute value 
of order parameters which are most stable in the case of $(\phi/2\pi, h)=(0, 0.2)$, $(0.25, 0.2)$, and $(0.25, 0.3)$, respectively. }
\label{FigLO}
\end{figure}
The gap equation is almost identical to Eq.~(\ref{Gap0}), except that we consider the finite temperature $T$ here: 
\begin{equation}
\Delta_i= g^2\sum_{\nu=1}^{2N}u_{i \uparrow}^\nu v_{i\downarrow}^{\nu\ast}
\mathrm{tanh}\frac{E_\nu}{2T}.
\label{GapN}
\end{equation}

The iterative calculations of Eqs.\ (\ref{BdGN}) and (\ref{GapN}) yield the pair potentials, eigenspinors, and eigenenergies.
In order to find the lowest energy state, we calculate the total free energy 
\begin{equation}
F=-T \sum_\nu \ln \left(1+e^{-E_\nu/T}\right)+\sum_i \frac{|\Delta_i|^2}{2g^2}
-\sum_{i}(\mu+h).
\end{equation} 
We set the chemical potential, the attractive potential, 
and the temperature to be $\mu=-0.5,~g=1.0,~T=0.005$, respectively, in the unit of  the transfer integral.  
Here the temperature is chosen to be much smaller than the critical temperature.
The size of the ring is $N=50$, which is supposed to be sufficiently large to reach the thermodynamic limit \cite{Quan}. 
We have prepared the 30 initial configurations and compared the free energies.
  
The ground states are categorized as follows.
If there is no amplitude modulation for the pair potential, 
the state is the BCS state or FF state, depending on if there is finite supercurrent (FF) or not (BCS). 
Here the supercurrent is defined by 
\begin{equation}
J=\frac{1}{2i}(\Delta^\ast \partial_x \Delta- \Delta\partial_x\Delta^\ast)-2\frac{\phi}{L} |\Delta|^2.
\label{scurrent}
\end{equation}
Here the last term on the right-hand side of Eq.\ (\ref{scurrent}) ensures the gauge invariance of supercurrent.
If the amplitude of the pair potential modulates, the states are again categorized by the supercurrent; 
if the supercurrent is zero, the state is the LO state and otherwise the FFLO state. 
\begin{figure}
\includegraphics[width=16pc]{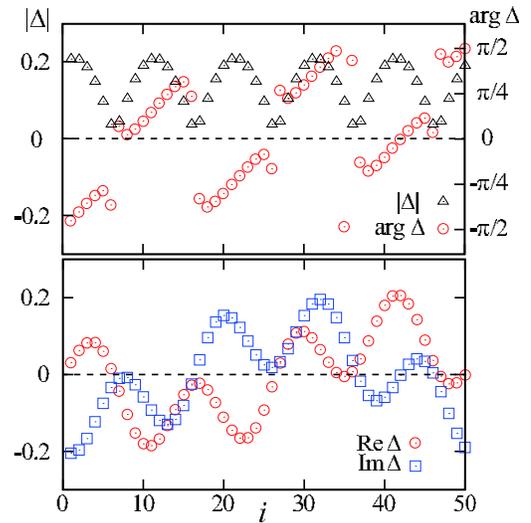}
\caption{The pair potentials of the FFLO phase for $\phi/2\pi=0.90,~h=0.35$. 
The upper figure shows that both the amplitude ($|\Delta|$) and phase (arg$\Delta$) of the pair potential  spatially modulate with the different periodicity. 
The lower figure shows that the real part of the pair potential and the imaginary part of the pair potential never touch to zero at the same point and thus $|\Delta|\neq 0$ for a whole region. }
\label{FFLOprofile}
\end{figure}
We can show that the LO state only appears in the absence of the supercurrent (see Appendix A). 
If there is no pair potential $\Delta=0$, the state is the normal state.  

The corresponding model has been already used in Refs.\ \cite{Quan} and \cite{Yoshida}. 
In Ref.\ \cite{Quan}, the phase transition between the LO phase and the FF phase was discussed. 
The existence of an additional phase called  the half vortex phase was mentioned in Ref.\ \cite{Yoshida}. 
The pair potential of the half vortex state proposed in Ref.\ \cite{Yoshida} has the form $\Delta\propto\cos(m\pi{x}/L)\times\exp({i\pi{nx}/L})$, with half integers $m,~n$. 
We show that this competing state between the LO and FF states are not the half vortex state but the FFLO state. 
We plot typical profiles of a FF state, LO states, and a FFLO state in Figs.\ \ref{FigFF}, \ref{FigLO}, and \ref{FFLOprofile}, respectively. 
In the case of the FFLO state, both the amplitude and the phase of the pair potential have spatial modulation. Moreover, the amplitude does not vanish in the whole region.  

\begin{figure}
\includegraphics[width=20pc]{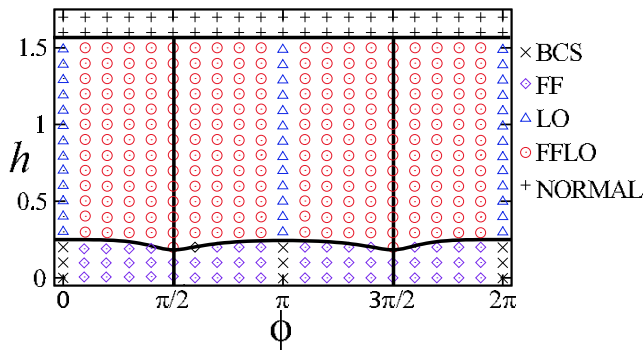}
\caption{Phase diagram in the $\phi-h$ plane. 
The FFLO phase is represented by the circle. 
The first-order transition lines are represented by the solid lines. 
The BCS phase and LO phase appears only on $\phi=n \pi$ ($n$ is integer) lines.
The phase diagram has periodicity $\pi$ in $\phi$ direction 
and reflection symmetry with respect to the $\phi=\pi/2$ line, which corresponds to the change in the direction of supercurrent. 
}
\label{phase_diagram}
\end{figure}
\begin{figure}
\includegraphics[width=20pc]{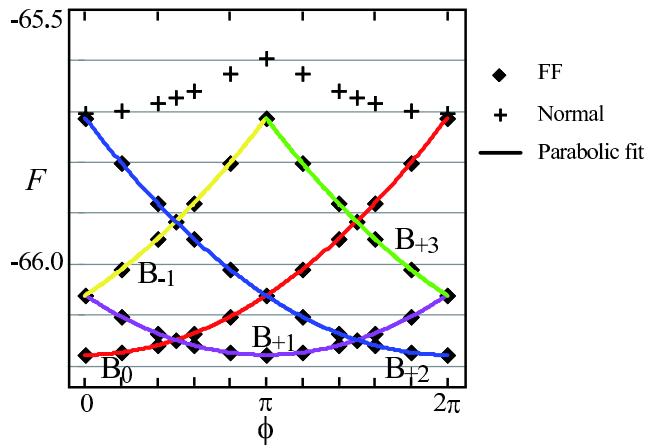}
\caption{The magnetic flux dependence of free energies in the case of $h=0$. 
The periodic structure appears in $\phi$ direction. 
The branch $\mathrm{B}_{+m}$ and $\mathrm{B}_{-n}$ has $+m$ or $-n$ 
flux, respectively, inside the ring relative to branch $\mathrm{B}_0$. 
The phase structures of FF states have $\pi/2$ periodicity,  
whereas the normal states have $\pi$ periodicity, 
which reflects the $2e$ charge of the condensates in FF states 
and $e$ charge for normal states, respectively.}
\label{FigFreeEnergyFF}
\end{figure}
\section{Phase diagram} 
Here we also show the phase diagram in Fig.\ \ref{phase_diagram} \cite{footnote2}. 
The magnetic fields are measured in the unit of the transfer integral. 
We show the first-order transition lines by the solid lines. 
We can see that the LO and FF phases are realized for $\phi=0$ and $h>0.2$, 
and for $h=0$ and $\phi\neq 0, \pi, 2\pi$, respectively. 
These results verify the naive discussion made above; 
the pair potential tends to rotate if the AB flux penetrates the ring, 
whereas it tends to have a spatial modulation in the presence of Zeeman field.

The most remarkable result is that the FFLO phase appears in a wide range of the parameters. 
In addition to this, we also see other characteristics. 
First, the phase diagram has periodicity $\pi$ in the $\phi$ direction. 
Second, the phase diagram has reflection symmetry with respect to $\phi=\pi/2$. 
Third, the BCS states and LO states also appear in the case of $\phi=n\pi$ with an arbitrary integer $n$. 
\begin{figure}
\includegraphics[width=16pc]{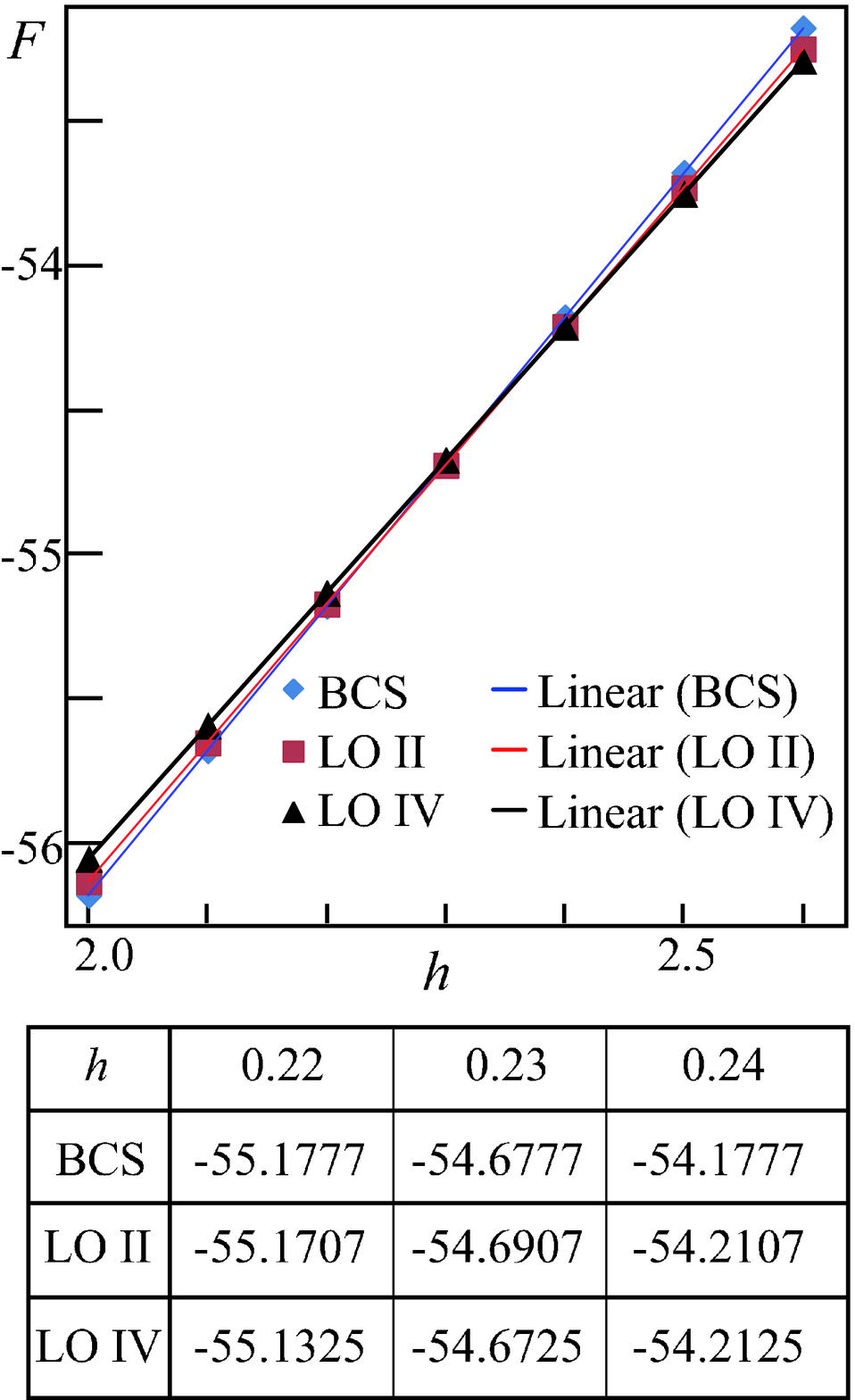}
\caption{The free energies as a function of $h$ for $\phi=0$. 
The LO II and LO IV are the LO phase with two peaks and four peaks for $|\Delta|$, respectively. 
The free energies have different slopes, and thus there are first-order transitions between BCS, LO II, and LO IV.}
\label{FigFreeEnergyLO}
\end{figure}
These facts can be understood by using $\hat\Delta(x)=e^{-2i\phi x/L}\Delta(x)$ defined in Sec.\ II. 
The boundary condition for $\hat \Delta$ becomes $\hat\Delta(x+L)=e^{2i\phi} \hat\Delta(x)$ as we have seen in Sec.\ II. 
By using $\hat \Delta$, the supercurrent (\ref{scurrent}) is rewritten as 
$J=(\hat \Delta^\ast \partial_x \hat \Delta-\hat\Delta \partial_x \hat\Delta^\ast)/2i$ \cite{footnote3}.
Thus the magnetic flux appears only in the boundary condition. 
This causes the periodic structure of the phase diagram in $\phi$, 
namely, the $\pi$ translation symmetry and reflection symmetry with respect to $\phi=\pi/2$. 
These are the consequences of the AB effect with an effective charge $2e$ for Cooper pairs. 
The change $\phi\leftrightarrow \pi-\phi$ corresponds to the change of the direction of supercurrent. 
The boundary condition for $\hat \Delta$ also suggests 
that the states without phase modulation can appear only in the case of $\phi=n\pi$ with an integer $n$. 

\begin{figure*}[t]
\includegraphics[width=30pc]{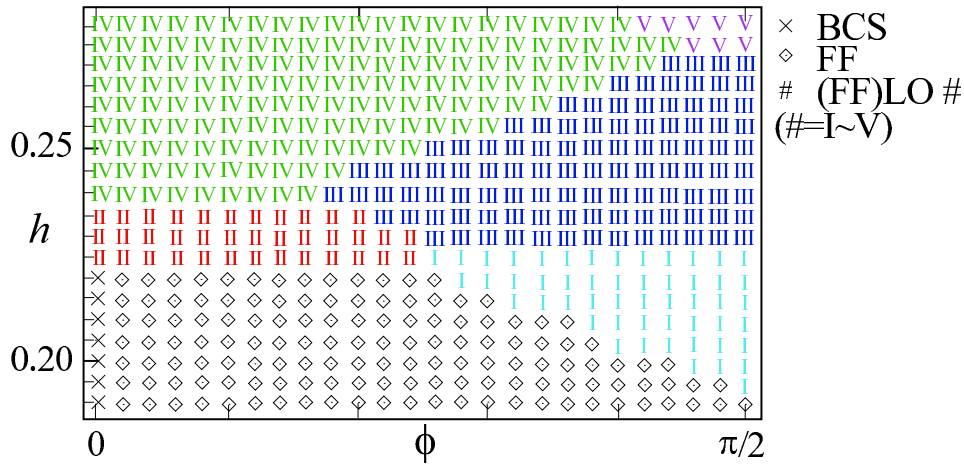}
\caption{The phase diagram near the phase boundary between BCS (FF) and LO (FFLO). 
Inside the LO or FFLO region in Fig.\ \ref{phase_diagram}, there are additional phase boundaries 
between the states with different number of nodes in $|\Delta|$. 
Here we classify states by the number of peaks in $|\Delta|$. 
The states I, II, and III correspond to those in Fig.\ \ref{FigLO}. 
}
\label{phase_diagram4}
\end{figure*}
In Fig.\ \ref{FigFreeEnergyFF}, we plot the free energy as a function of $\phi$ in the case of $h=0$. 
This clearly shows the periodic structure of the FF states. 
This phenomenon is known as the Little-Parks effect \cite{Little-Parks}. 
The branch $\mathrm{B}_{+m}$ and $\mathrm{B}_{-n}$ has $+m\pi$ or $-n\pi$ flux, respectively,  inside the ring relative to branch $\mathrm{B}_0$. 
At $\phi=\pi/2$ (or equivalently, the number of the vortex is $2\phi/2\pi=1/2$), 
two states degenerate: the state with supercurrent flowing to the positive direction in branch $\mathrm{B}_0$ and that with the supercurrent flowing to the opposite direction in branch $\mathrm{B}_{+1}$. 
If we increase the flux from $\phi=0$ to $\phi=\pi/2+0$, then one more quantum of flux is captured in the ring  (phase slip) and the direction of the supercurrent changes. 
Thus the first-order transition occurs at $\phi=\pi/2$. 
This first-order transition line continues to the boundary of FFLO and the normal state. 
We can also see the reflection symmetry with respect to $\phi=\pi$ for the normal states. 
This period is twice larger than that of the FF states. 
This difference comes from the fact that the charge is not $2e$ but $e$ for the particles in normal states. 

In Fig.\ \ref{phase_diagram}, we present the first-order transition lines separating BCS (FF) states and (FF)LO states and those separating (FF)LO states and normal states. 
In addition to these first-order transition lines phase, we find several first-order transition lines inside the (FF)LO regions. 
In Fig.\ \ref{FigLO}, we plot the magnitude of the order parameter for LO or FFLO states for various parameters. 
The labels I, II, and III correspond to the number of the peaks in $|\Delta|$. 
We plot the free energies as a function of $h$ in Fig.\ \ref{FigFreeEnergyLO}, 
and we show the value of the free energies near the transition point in the table. 
The slopes of the free energies are different from each other, 
clearly showing the first-order phase transitions.

In Fig.\ \ref{phase_diagram4}, we show the phase diagram 
that separates the (FF)LO states with different number 
of peaks by the first-order transitions. 
This rich phase structure stems from the existence of two contributions to the free energy 
that compete with each other: the energy of the excess spins and the condensation energy.
 
It is shown that only the states with an even number of peaks in $|\Delta|$ 
appear for $\phi=0$, whereas only the states with an odd number of peaks in $|\Delta|$ appear for $\phi=\pi/2$. 
In the intermediate region of $0\le \phi\le \pi/2$, 
there is an odd-even phase transition. 
As an example, if we increase $h$ from $h=0$ for $\phi=2\pi/5$, the number of the peaks changes to 1, 3, 4 in the order.

\section {Summary and discussion} 
In conclusion, 
we generalized the method to obtain the analytical solutions for the BdG equation and the gap equation 
to the case with magnetic fields. 
By using this method, we showed that the novel FFLO solutions proposed recently 
are also the solutions for the superconducting ring threaded by the AB flux and with the Zeeman field on the ring. 
We have demonstrated that this FFLO phase can be realized 
as the lowest energy state of our system instead of the half vortex state. 
We have shown the phase diagram as a function of the AB flux and the Zeeman field. 
The FFLO states, which enjoy an analytical description involving only a few parameters, together with the excess spin contribution result in a nontrivial phase structure.

While our phase diagram is calculated by taking the number of sites equal to 50, we have also confirmed that the configurations remain qualitatively 
the same even if we increase the number of sites to 59, 73, and 100 for some choices of the parameters. 
This supports the fact that our results are free from finite-size artifacts. 

In Refs.~[\onlinecite{Takahashietal}] and [\onlinecite{Takahashinitta}], exact self-consistent solutions were found in quasi-1D, in which
twisted kinks with arbitrary phase shifts are separated at arbitrary distances. 
A ring version of this case may be stabilized by nonuniform magnetic fields.

It was shown in Ref.~[\onlinecite{Hofmann:2010gc}] that 
the Gross-Neveu model in 1+1 dimensions 
appears as a low-energy effective theory in 3+1 dimensions under a strong uniform  magnetic field. 
Therefore our setup may be realized in a certain region of the QCD phase diagram \cite {Casalbuoni, Anglani}. 

\section*{Acknowledgments}
We thank H. Watanabe for fruitful discussion. R.\ Y.\ also appreciates P.\ Bruno for his critical comments. 
R.~Y.~was a research fellow supported by a Grant-in-Aid for Scientific Research (S) (No. 26220711) from the Japan Society for the Promotion of Science, 
and the work of R.~Y.~is also partially supported by the Yukawa Memorial Foundation. 
The work of R.\ Y., S.\ T., and H.\ H.\ is partially supported by a Grant-in-Aid for Scientific Research 
(No.\ 25287098). 
The work of M.~N.~is supported in part by a Grant-in-Aid for
 Scientific Research on Innovative Areas ``Topological Materials
 Science'' (KAKENHI Grant No.~15H05855) and ``Nuclear Matter in Neutron
 Stars Investigated by Experiments and Astronomical Observations''
 (KAKENHI Grant No.~15H00841), from the Ministry of Education,
 Culture, Sports, Science (MEXT) of Japan. The work of M.~N.~is also
 supported in part by a Japan Society for the Promotion of Science
 (JSPS) Grant-in-Aid for Scientific Research (KAKENHI Grant
 No.~25400268) and by the MEXT-Supported Program for the Strategic
 Research Foundation at Private Universities, ``Topological Science''
 (Grant No.~S1511006).

\appendix
\section{Basic properties of  order parameters}
Here we briefly summarize the basic properties of the order parameters. 
One can derive the non linear Schr\"odinger equation (NLSE) from the BdG equation and the gap equation as 
\begin{equation}
\hat\Delta^{\prime\prime}+i(b-2E)\hat\Delta^\prime-2(a-Eb)\hat\Delta-2\hat\Delta|\hat\Delta|^2=0.
\label{NLSE}
\end{equation}
All the solutions known before, including BCS, FF, LO, and the twisted kink crystal state, obey the above equation with suitable real parameters $a$, $b$. 

First, we can show that the divergence of the supercurrent can be calculated as
\begin{equation}
\frac{1}{2i}(\hat\Delta^\ast\hat\Delta^\prime-\hat\Delta^{\ast\prime}\hat\Delta)^\prime
=-\frac{b-2E}{2}(|\hat\Delta|^2)^\prime.
\end{equation}
This equality is easily verified by calculating $\text{NLSE}\times \hat\Delta^\ast- (\text{NLSE})^\ast \times \hat\Delta$.

Second, we can show that $b=2E$ for the LO phase. 
In the case of the LO phase, the order parameter can be written as 
$\hat\Delta=f(x) e^{i\delta}$, with the real function $f(x)$ and a constant $\delta$. 
Thus the NLSE becomes 
\begin{equation}
f^{\prime\prime}+i(b-2E)f^\prime-2(a-Eb)f-2f^3=0.
\end{equation}
The imaginary part of the above equation leads to
\begin{equation}
i(b-2E)f^\prime=0.
\end{equation}
This equation requires $b$ to be $2E$ for the LO phase. 

We can also show that if $b=2E$, the order parameter $\hat\Delta=f(x)e^{i\delta(x)}$ with real functions $f(x)$ and $\delta(x)$ must satisfy that $\delta(x)$ is constant or $f(x)=0$. 
By substituting $\hat\Delta=f(x)e^{i\delta(x)}$ into NLSE with $b=2E$, we obtain
\begin{equation}
f^{\prime\prime}-\delta^{\prime 2}f +2i\delta^\prime f -2(a-Eb)f-2f^2=0.
\end{equation}
Thus the imaginary part of the above equation requires $\delta^\prime =0$ or $f=0$.
The contraposition of this shows that if $f(x)\neq 0$ nor $\delta^\prime(x)\neq0$, $b$ must not be $2E$.


\begin{thebibliography}{99}
\bibitem{FF} P.\ Fulde and R.\ A.\ Ferrell, Phys.\ Rev.\ {\bf 135}, A550 (1964).

\bibitem{LO} A.\ I.\ Larkin and Y.\ N.\ Ovchinnikov,
	Zh.\ Eksp.\ Teor.\ Fiz.\ {47}, 1136 (1964) [Sov.\ Phys.\ JETP {\bf 20},
	762 (1965).]

\bibitem{Machida} K.\ Machida and H.\ Nakanishi, Phys.\ Rev.\ B \textbf{30}, 122 (1984).

\bibitem{Aoyama}K.\ Aoyama, R.\ Beaird, D.\ E. Sheehy, and I.\ Vekhter, Phys.\ Rev.\ Lett. {\bf 110}, 177004 (2013).
\bibitem{Klein} U.\ Klein, Phys.\ Rev.\ B\ {\bf 69}, 134518 (2004).
\bibitem{Zyuzin} A.\ A.\ Zyuzin and A.\ Y.\ Zyuzin, Phys.\ Rev.\ B {\bf 79}, 174514 (2009).


\bibitem{Buzdin3} A.\ I.\ Buzdin, Rev.\ Mod.\ Phys.\ {\bf 77}, 935 (2005).
\bibitem{Bergeret}F.\ S.\ Bergeret, A.\ F.\ Volkov, and K.\ B.\ Efetov, 
Rev.\ Mod.\ Phys.\ {\bf 77}, 1321 (2005). 

\bibitem{Radzihovsky} L.\ Radzihovsky and D.\ E.\ Sheehy, Rep.\ Prog.\ Phys.\ {\bf 73}, 076501
(2010); L.\ Radzihovsky, Phys.\ Rev.\ A {\bf 84}, 023611 (2011).

\bibitem{Liao} Y.\ Liao, A.\ Sophie, C.\ Rittner, T.\ Paprotta,	W.\ Li, G.\ B.\ Partridge,	R.\ G.\ Hulet, S.\ K.\ Baur, and E.\ J.\ Mueller, Nature (London) {\bf 467}, 567 (2010).


\bibitem{Casalbuoni} 
R.\ Casalbuoni and G.\ Nardulli, Rev.\ Mod.\ Phys.\ {\bf 76}, 263 (2004).

\bibitem{Anglani} 
R.\ Anglani, R.\ Casalbuoni, M.\ Ciminale, N.\ Ippolito, R.\ Gatto, M.\ Mannarelli, and M.\ Ruggieri, Rev.\ Mod.\ Phys.\ {\bf 86}, 509 (2014).


\bibitem{Radovan} H.\ A.\ Radovan et al., Nature (London) {\bf 425}, 51 (2003); 
Y.\ Matsuda and H. Shimahara, J.\ Phys.\ Soc.\ Jpn.\ {\bf 76}, 051005 (2007); 
Y.\ Yanase and M.\ Sigrist, ibid.\ {\bf 78}, 114715 (2009). 
\bibitem{Yonezawa} S. Yonezawa, S. Kusaba, Y. Maeno, P. Auban-Senzier, C. Pasquier,
K. Bechgaard, and D. Jerome, Phys. Rev. Lett. 100, 117002 (2008);
A. G. Lebed and S. Wu, Phys. Rev. B 82, 172504 (2010).
\bibitem{Wu} S. Wu and A. G. Lebed, Phys.\ Rev.\ B {\bf 80} 035128 (2009); S. Brazovskii,
J.\ Supercond.\ Novel Magnetism {\bf 20}, 489 (2007).


\bibitem{Quan} H.\ T.\ Quan and J.\ -X.\ Zhu, Phys.\ Rev.\ B {\bf 81}, 014518 (2010).
\bibitem{Yoshida} T.\ Yoshida and Y.\ Yanase, Phys.\ Rev.\ A {\bf 84}, 063605 (2011).

\bibitem{Basar} G.\ Ba\c{s}ar and G.\ V.\ Dunne, Phys.\ Rev.\ Lett.\ {\bf 100},
	200404 (2008).
\bibitem{Basar2} G.\ Ba\c{s}ar and G.\ V.\ Dunne, Phys.\ Rev.\ D
	{\bf 78}, 065022 (2008).

\bibitem{Basar:2009fg} 
  G.~Basar, G.~V.~Dunne, and M.~Thies, Phys.\ Rev.\ D {\bf 79}, 105012 (2009).

\bibitem{BdG} P. G. de Gennes, {\it Superconductivity of Metals and Alloys}
(Benjamin, New York, 1966).

\bibitem{Barsagi} J.\ Bar-Sagi and C.\ G.\ Kuper, Phys.\ Rev.\ Lett.\ {\bf 28}, 1556 (1972).

\bibitem{YoshiiPRB} R.\ Yoshii, S.\ Tsuchiya, G.\ Marmorini, and M.\ Nitta, Phys.\ Rev.\ B {\bf 84}, 024503 (2011). 
\bibitem{YoshiiJPSJ}R.\ Yoshii, G.\ Marmorini, and M.\ Nitta, J.\ Phys.\ Soc.\ Jpn.\ {\bf 81}, 094704 (2012).

\bibitem{Shei}S.\ S.\ Shei, Phys.\ Rev.\ D {\bf 14} (1976) 535.

\bibitem{footnote2}
{The phase diagram is periodic in $\phi$ with periodicity $2\pi$. Thus we restrict the range of the phase diagram as shown in Fig.\ \ref{phase_diagram}. It should be noted that the free energy (12) allows a solution with higher flux, even for $|\phi|>\pi$, but this state is not the lowest energy state, which can decay through a phase slip.} 

\bibitem{footnote3}
{Here we note that this is not the gauge transformation but the extraction of the phase from $\Delta$. Thus the form of the supercurrent is changed.}

\bibitem{Little-Parks} 
W.\ A.\ Little and R.\ D.\ Parks, Phy.\ Rev.\ Lett.\ {\bf 9}, 9 (1962).

\bibitem{Takahashietal} D.\ A.\ Takahashi, S.\ Tsuchiya, R.\ Yoshii, and M.\ Nitta, Phys.\ Lett.\ B {\bf 718}, 632 (2012).
\bibitem{Takahashinitta} D.\ A.\ Takahashi and M.\ Nitta, Phys.\ Rev.\ Lett.\ {\bf 110}, 131601 (2013).

\bibitem{Hofmann:2010gc} 
  J.~Hofmann, Phys.\ Rev.\ D {\bf 82}, 125027 (2010).






\end{thebibliography}
\end{document}